\documentclass[aps,prb,twocolumn,groupedaddress,showpacs]{revtex4}

\usepackage{graphics}

\begin{document}

\preprint{NSF-KITP-03-24}

\title{Shot noise of spin polarized electrons}

\author{A. Lamacraft}
\affiliation{Department of Physics, Princeton
University, Princeton, NJ, 08544, USA. }
\date{September 21, 2003}
\email{alamacra@princeton.edu}

\begin{abstract}

The shot noise of spin polarized electrons is shown to be generically dependent upon spin-flip processes. Such a situation represents perhaps the simplest instance where the two-particle character of current fluctuations out of equilibrium is explicit, leading to trinomial statistics of charge transfer in a single channel model. We calculate the effect of spin-orbit coupling, magnetic impurities, and precession in an external magnetic field on the noise in the experimentally relevant cases of diffusive wires and lateral semiconductor dots, finding dramatic enhancements of the Fano factor. The possibility of using the shot noise to measure the spin-relaxation time in an open mesoscopic system is raised.
\end{abstract}


\maketitle

Quantum statistical fluctuations of current and voltage in mesoscopic conductors encode information beyond that obtainable from averaged measurements. The second moment of current contains two-particle correlations as subtle in principle as the one-particle signatures of phase coherence - weak localization, universal conductance fluctuations, and so on -  revealed by conductance measurements. The full counting statistics (FCS)~\cite{Levitov,FCSbook} - the probability distribution of charge transfer through a conductor - similarly depends on the multiparticle correlations of a many electron system. From a theoretical standpoint the simplest conductor is a scatterer in a one dimensional channel of non-interacting fermions. Here, the FCS are binomial with a number of attempts $eV\tau/h$, where V is the voltage bias and $\tau$ the observation time, and a success probability for charge transfer $T$, the transmission coefficient of the scatterer. Thus, although the existence of a finite attempt frequency is a signature of the correlations imposed by the exclusion principle, the `elementary events' are still described by one-particle probabilities.

It is the purpose of this paper to describe the simplest situation in which multiparticle effects enter in a more essential way, leading to particular signatures in the second moment and to a novel \textit{trinomial} statistics of charge transfer. We will first describe the physics in an idealized conductor, before moving on to the more realistic cases of diffusive wires and semiconductor quantum dots. 

We will be concerned with the injection of spin-polarized currents into a scatterer where spin-flip processes may occur. The basic phenomenon was identified in a recent Letter~\cite{ebl} for a particular scattering geometry and spin-orbit interaction. In fact, the shot noise of spin-polarized electrons is generically dependent on spin-flip processes, as these determine the triplet to singlet scattering amplitude involved in scattering electrons from two different channels into the same channel. We predict dramatic enhancements of the noise-to-current ratio (Fano factor) as the injected spin relaxes, raising the possibility of using the noise to measure spin-relaxation rates. It is remarkable that a charge transport measurement generically contains information on the dynamics of spin. The effect of spin flips on shot noise for particular spin-valve geometries was also discussed in two recent preprints~\cite{mish,bz}

\begin{figure} 
\begin{center}
\setlength{\unitlength}{2.8in}
\begin{picture}(1, 0.75)(0,0)
  \put(0,0){\resizebox{1\unitlength}{!}{\includegraphics{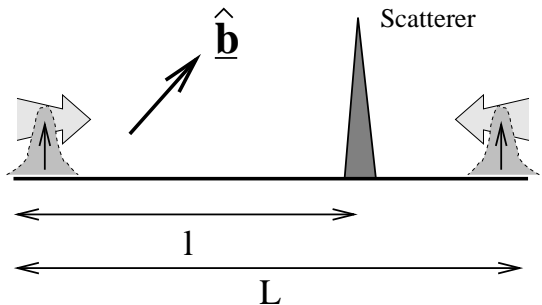}}}
  \end{picture}
\end{center}
\caption{One-dimensional model illustrating the effect of spin-precession on two-particle scattering. Spin up electrons enter from left and right, scattering from a potential scatterer while precessing in an external magnetic field.
\label{fig:wire}}
\end{figure}

Consider first the two-particle scattering problem illustrated in Fig.~\ref{fig:wire}. 
Spin polarized electrons enter a one-dimensional channel of length $L$ from left and right. The Hamiltonian is given by  $\hat H=\frac{\hbar^2 k^2}{2m^*}+(\epsilon_Z/2) \hat\mathbf{b}\cdot\mathbf{\sigma}$, where the unit vector $\hat\mathbf{b}$ gives the direction of the external magnetic field, and $\epsilon_Z$ is the Zeeman splitting.  In the channel at distance  $\ell$  from one end is a potential scatterer with S-matrix 
\[{\cal S}_{\mathrm{scat}}=\pmatrix{r & t'  \cr t & r'}_{LR}\otimes\openone_s\;,\]
with $\openone_s$ indicating the identity in the spin sector. The overall S-matrix of the structure, linearizing the dispersion near the fermi energy,  is thus 
\begin{eqnarray*}
&&{\cal S}=\pmatrix{rU(2\ell) & t'\; U(L) \cr t\; U(L) & r'U(2(L-\ell))}_{LR}\\
&&U(x)=\exp\left(-i\frac{\epsilon_Z x}{2 v_F}\hat\mathbf{b}\cdot\mathbf{\sigma}\right)\;,
\end{eqnarray*}
where $v_F$ is the Fermi velocity. The angle of precession is determined by the length of time the electrons spend in the channel. There are three final outcomes for the location of the two electrons: both leave to the left, both to the right, or one in each direction. Accounting for fermi statistics, and assuming that the electrons have the same energy, these probabilities are.
\begin{eqnarray}\label{2probs}
\text{Both to left or right:} &&TR|U_{\downarrow\uparrow}(L-2\ell)|^2\\\nonumber
\text{One in each direction:} &&T^2+R^2+2TR|U_{\uparrow\uparrow}(L-2\ell)|^2 \;,
\end{eqnarray}
where $T=|t|^2=|t'|^2$, $R=1-T$. We see that only through the spin rotation are both electrons allowed to leave to the same contact. They must form a singlet: $TR|U_{\downarrow\uparrow}(L-2\ell)|^2$ is the probability for this event. If the scatterer lies precisely in the middle of the channel ($L=2\ell$), reflected and transmitted electrons precess through the same angle, so the probability to form a singlet is zero. Note that assuming that electrons entering the channel are polarized in a given direction is equivalent to having an inhomogenous magnetic field, vanishing outside the channel. A uniform field conserves total spin so that there is no triplet to singlet amplitude. Scattering due to band distortions in the inhomogenous field can be neglected if we assume the scale of variation is much longer than the Fermi wavelength.

The implication of this consideration for the statistics of transmitted charge is the following. Evidently no charge is transmitted on average, but for $|U_{\downarrow\uparrow}(L-2\ell)|^2\neq 0$, there will be uncertainty in the amount of charge that has passed through the system, as there is a non-zero probability to transport two charges in either direction. To make this connection more formal, let us compute the FCS for this problem. We obtain the generating function $\chi(\lambda, \tau)=\sum_n P(n,\tau) e^{i\lambda n}$ of the probabilities $P(n,\tau)$ to pass $n$ charges in time $\tau$ using the result~\cite{Levitov}
\begin{eqnarray*}
&&\ln[\chi(\lambda,\tau)]=\frac{\tau}{2\pi\hbar}\int dE \;\mathrm{tr}\,\ln\left(1+\hat n(E)({\cal S}^{-1}_{-\lambda}{\cal S}_{\lambda}-\openone)\right)\\
&&{\cal S}_{\lambda}=\pmatrix{e^{i\lambda/2} & 0 \cr 0 & e^{-i\lambda/2}}_{LR}{\cal S}\pmatrix{e^{-i\lambda/2} & 0 \cr 0 & e^{i\lambda/2}}_{LR}\;,
\end{eqnarray*}
where $\hat n(E)=\mathrm{diag}(n_{L\,\uparrow},n_{L\,\downarrow},n_{R\,\uparrow},n_{R\,\downarrow})$ is the distribution function of the incoming electrons. The injection of an excess of spin up electrons into both ends of the channel corresponds to $n_{L,R\,\uparrow,\downarrow}(E)=\theta(\mu_{\uparrow,\downarrow}-E)$ at zero temperature, with $\Delta\mu\equiv\mu_{\uparrow}-\mu_{\downarrow}>0$. We find
\begin{eqnarray*}
\chi(\lambda,\tau)&=&\left(T^2+R^2+2TR|U_{\uparrow\uparrow}(L-2\ell)|^2\right.\\*
&&\left.+2TR|U_{\downarrow\uparrow}(L-2\ell)|^2\cos 2\lambda\right)^{\Delta\mu\tau/h}\;.
\end{eqnarray*}
This is the generating function of a trinomial distribution with number of attempts $\Delta\mu\tau/h$ and probabilities to transfer $\pm 2$ and $0$ charges given by Eq.~\ref{2probs}. The corresponding noise is~\cite{foot1} 
\[S\equiv\lim_{\tau\to\infty}e^2\langle\Delta n^2\rangle/\tau=\frac{8e^2}{h}\Delta\mu TR|U_{\downarrow\uparrow}(L-2\ell)|^2\;,\]
which displays a dependence on the spin flip probability $|U_{\downarrow\uparrow}|^2$ that mirrors the result of  Ref.~\onlinecite{ebl} in a different geometry, where the spin rotation was due to spin-orbit coupling. Generally, a non-trivial dependence of the noise on parameters governing spin-flip scattering may be expected when there are two or more channels of incoming spin polarized electrons, for then the noise will contain a contribution corresponding to two electrons in a triplet state and different incoming channels passing to a singlet in the same outgoing channel. Spin flips may be caused by magnetic impurities, spin-orbit scatterers, or precession in an external magnetic field if contributing trajectories are of different lengths, as in the above example. In the remainder of this paper we will discuss the implications of this observation for noise in diffusive wires and semiconductor quantum dots, where the corresponding measurements in the unpolarized case have been made~\cite{henny,Oberholzer}.

%
%

In such mesoscopic scatterers, an average over scattering matrices is called for. Note that Lesovik's~\cite{lesovik} formula $S=\frac{e^2}{h}\Delta\mu\sum_n T_n(1-T_n)$, expressing the noise in terms of the eigenvalues $T_n$ of the transmission matrix $t^{\dagger}t$, does not apply in the present situation: the density matrix is non-trivial in the incoming channels, so that a rotation to the block diagonal S-matrix is not possible.  Thus one route to the calculation of the averaged noise - averaging over the distribution of $T_n$~\cite{bb} - is not available. 
\begin{figure} 
\begin{center}
\setlength{\unitlength}{2.8in}
\begin{picture}(1, 0.75)(0,0)
  \put(0,0){\resizebox{1\unitlength}{!}{\includegraphics{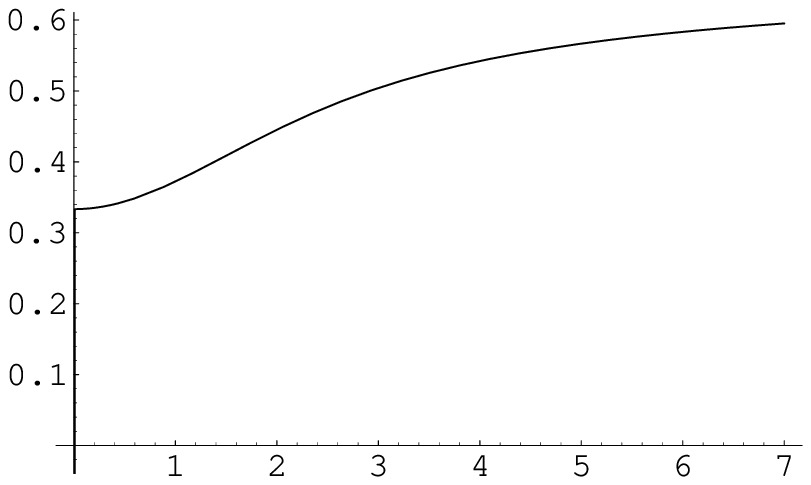}}}
  \put(0.35,0.15){\resizebox{0.7\unitlength}{!}{\includegraphics{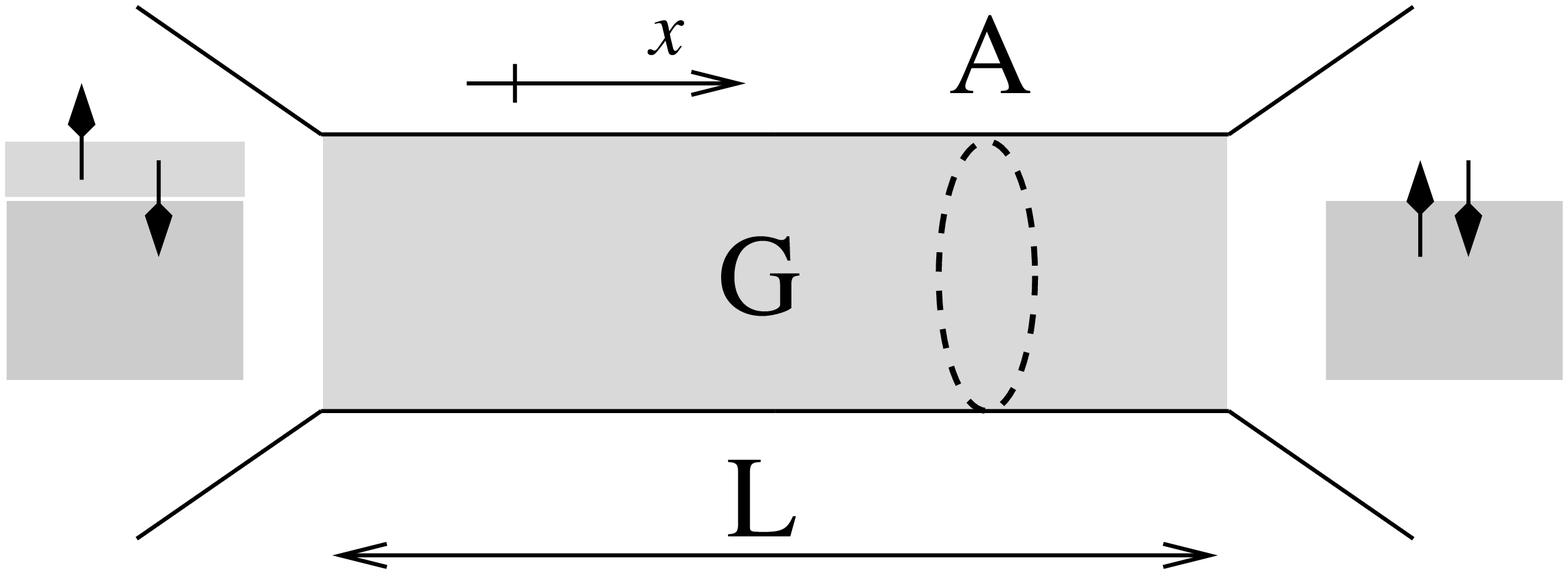}}}
  \put(0.54,-0.05){\makebox(0,0)[b]{$L/L_s$}}
  \put(-0.05,0.35){\rotatebox{90}{\makebox(0,0)[t]{$F$}}}
\end{picture}
\end{center}
\caption{Fano factor (defined by $S=FeI$) for a diffusive wire passing a spin-polarized current, as a function of spin-relaxation length $L_s$. Inset: experimental geometry
\label{fig:Fano}}
\end{figure}

\emph{Diffusive wires}. We consider the geometry depicted in the inset to Fig~\ref{fig:Fano}, a diffusive wire of length $L$ and cross-sectional area $A$, with diffusion constant $D$. A microscopic calculation of the shot noise of a diffusive wire was presented in Ref~\onlinecite{aly}. Here we pursue a formally equivalent approach based on Nazarov's circuit theory~\cite{nazarov}. Current correlators may be obtained from the Keldysh Green's function satisfying
\[ \left(i\partial_t-\hat H-\tau_3\frac{\lambda}{2L}\hat J_x\right)\hat G_{\lambda}(\mathbf{r},\mathbf{r'}t,t')=\delta(\mathbf{r}-\mathbf{r'})\delta(t-t')\;,\]
where $\tau_i$ denote the Pauli matrices in Keldysh space, and $\hat J_x=-ie(\overrightarrow{\partial_x}-\overleftarrow{\partial_x})/2m$ is the $x$-component of the one-particle current density operator . With this definition we can write the generating function~\cite{Levitov,nazarov}
\begin{eqnarray*} 
\chi(\lambda,\tau)&=&\langle {\cal T}_K \exp\left(\frac{i\lambda}{2L}\int_0^{\tau} dt \int d\mathbf{r}\, \hat\Psi^{\dagger}\tau_3\hat J_x\hat\Psi\right)\rangle\\*
&=& \exp\left(-\frac{1}{2} \int_0^{\lambda} d\lambda'\,\mathrm{tr}\{\tau_3\hat J_x\hat G_{\lambda'} \}\right) \;,
\end{eqnarray*}
where the trace is understood to be over the Keldysh and spin spaces, as well as space and time indices. ${\cal T}_K$ denotes time ordering along the Keldysh contour. In the above we depart slightly from the usual formulation in that we take the `counting field' $\lambda$ to be constant throughout the conductor instead of defining a surface at which we measure the current. This does not affect the result as the zero-frequency correlator $\langle I(x)I(x')\rangle$ is independent of $x$ and $x'$ by current conservation. Using standard quasiclassical methods in the diffusion approximation we arrive at the following formulation of the disorder averaged problem (after Keldysh rotation)\cite{lo}
\begin{eqnarray} \label{usadel}
\langle\mathrm{tr}\{\tau_3 \hat J_x\hat G_{\lambda}\} \rangle_{\mathrm{dis}}&\to&-\frac{ie\nu D\tau}{2}\int dE\, \mathrm{tr}'\{\tau_1\hat g_{\lambda\,E}\partial_x\hat g_{\lambda\,E}\}\nonumber\\*
D\tilde\nabla(\hat g_{\lambda\,E}\tilde\nabla \hat g_{\lambda\,E})&-&i\epsilon_Z/2[\hat\mathbf{b}\cdot\mathbf{\sigma},\hat g_{\lambda\,E}\nonumber]\\*
&-&\frac{1}{2\tau_s}[\sigma_i\hat g_{\lambda\,E}\,\sigma_i,\hat g_{\lambda\,E}]=0\;,
\end{eqnarray}
where $D$ is the diffusion constant, $\nu$ is the density of states at the Fermi energy, $\hat g(\mathbf{r})=i\hat G(\mathbf{r},\mathbf{r})/\pi\nu$ is the quasiclassical Green's function, and $\tilde\nabla=\nabla-i\lambda/2L[\tau_3,\cdots]$. $\mathrm{tr}'$ denotes a trace over the Keldysh and spin spaces only. We have introduced the total spin relaxation rate $1/\tau_s$ due to spin-orbit and magnetic impurities. To find the noise we require a solution of Eq.~\ref{usadel} to first order in $\lambda$ only. The zeroth order solution has the usual form
\[\hat g_E^{(0)}(\mathbf{r})=\pmatrix{ \openone & 2{\cal F}_E(\mathbf{r}) \cr 0 & -\openone}\;,\]
where  ${\cal F}_E$ is a $2\times 2$ matrix in spin space, related to the distribution function ${\cal N}_E$ by ${\cal F}_E\equiv \openone-2{\cal N}_E$. We chose the simplest model for spin injection from the left: a half-metallic ferromagnet connected to the wire through a perfect interface, providing a reservoir of spin up electrons only at chemical potential $\mu_{\uparrow}$. At the right we have a normal reservoir with no spin polarization and chemical potential $\mu_N$. Since we will derive a general expression for the noise in terms of ${\cal F}_E$, extending the calculations to more realistic injection scenarios involving incomplete polarisation and interface resistances is a matter of kinetics. With the decomposition ${\cal F}_E=F_{E\,0}\openone+\mathbf{F}_E\cdot\mathbf{\sigma}$, the corresponding boundary conditions are
\begin{eqnarray} \label{bc}
&&F_{E\,0}(-L/2)=1/2\left[F_{\mathrm{eqm}}(E-\mu_{\uparrow})+F_{\mathrm{eqm}}(E-\mu_N)\right]\nonumber\\
&&F_{E\,0}(L/2)=F_{\mathrm{eqm}}(E-\mu_N)\nonumber\\
&&\mathbf{F}_E(-L/2)=1/2\left[F_{\mathrm{eqm}}(E-\mu_{\uparrow})-F_{\mathrm{eqm}}(E-\mu_N)\right]\hat\mathbf{s}\nonumber\\
&&\mathbf{F}_E(L/2)=0\;,
\end{eqnarray}
where $F_{\mathrm{eqm}}(E)=\tanh(E/2kT)$ corresponds to the Fermi distribution. The singlet and triplet parts of the distribution function satisfy
\begin{eqnarray*}
&&D\nabla^2 F_{E\,0}=0\\
&&D\nabla^2 \mathbf{F}_E-\epsilon_Z\hat\mathbf{b}\times\mathbf{F}_E-\frac{2}{\tau_s}\mathbf{F}_E=0\;.
\end{eqnarray*}
The solution is simplest when the Zeeman term is absent. 
\begin{eqnarray*}
F_{E\,0}(x)=(1/2-x/L)F_{E\,0}(-L/2)+(1/2+x/L)F_{E\,0}(L/2)\\
\mathbf{F}_E(x)=\frac{\mathbf{F}_E(-L/2)}{1-e^{-2L/L_s}}\left(e^{-(x+L/2)/L_s}-e^{(x-3L/2)/L_s}\right)\;,
\end{eqnarray*}
where $L_s=\sqrt{D\tau_s/2}$. The zeroth order result is thus $\mathrm{tr}'\{\tau_1\hat g_E\partial_x \hat g_E\}^{(0)}=4F_{E\,0}'$, and involves only the singlet part: the average current is unaffected by spin polarization. Substituting into Eq.~\ref{usadel} gives the first order correction
\begin{eqnarray} \label{int}
\mathrm{tr}'\{\tau_1\hat g_E\partial_x \hat g_E\}^{(1)}=
4i\lambda(F_{E\,0}^2+\mathbf{F}_E^2-1)\;.
\end{eqnarray}
The constant of integration in Eq.~\ref{int} is fixed by the fluctuation-dissipation theorem in equilibrium. We thus obtain $\ln[\chi(\lambda)]/\tau=i\lambda I-\lambda^2S/2+\cdots$, where $I=G\Delta\mu/2e$ is the average current ($G=2e^2\nu DA/L$ is the conductance including spin degeneracy) and the noise is
\begin{equation} \label{diff}
S=\frac{G}{2L}\int^{L/2}_{-L/2} dx \int dE \left[1-F_{E\,0}^2(x)-\mathbf{F}^2_E(x)\right]\;,
\end{equation}
Performing the $x$ and $E$ integrals in Eq.~\ref{diff} gives
\begin{widetext}
\begin{eqnarray*}
S=G\left(\frac{1}{3}\left[4kT+\Delta\mu\coth\left(\frac{\Delta\mu}{2kT}\right)\right]+\frac{(L_s/L)(1-e^{-4L/L_s})-4e^{-2L/L_s}}{2(1-e^{-2L/L_s})^2}\left[kT-(\Delta\mu/2)\coth\left(\frac{\Delta\mu}{2kT}\right)\right]\right)\;.
\end{eqnarray*}
\end{widetext}
One may verify that this result satisfies the following limits. As $\Delta\mu\equiv\mu_{\uparrow}-\mu_N\to 0$, the second term vanishes (originating from the triplet part) and we are left with $S=2GkT$, as required by the fluctuation-dissipation theorem. As $L_s\to\infty$ and at $T=0$, the noise is $S=(1/3)eI$. This is the familiar result for unpolarized electrons~\cite{bb}, indicating that without spin relaxation there is no signature of spin polarization in the noise. As $L_s$ decreases from infinity to zero, the Fano factor - defined by $S=FeI$ - changes from $1/3$ to $2/3$ (see Fig.~\ref{fig:Fano}). Note that although we used a microscopic quantum mechanical approach, the apperarance in Eq.~\ref{diff} of only quantities from kinetic theory indicates that phase coherence is not necessary to observe these effects~\cite{nagaev}

\emph{Quantum dots}. Spin-dependent transport phenomena have recently been the subject of intense investigation in lateral semiconductor quantum dots~\cite{af,bch,zum,mil}. Again, relaxation of injected spin-polarized current may be due to spin-orbit coupling or precession in an  external field. As in the case of the diffusive wire, we will  consider the shot noise in a two terminal set-up where a spin-polarized current is driven through one of the terminals. In the spin degenerate case, we have the well-known result $F=N_LN_R/N^2$, valid when $N\equiv N_L+N_R\gg 1$, where $N_L$ and $N_R$ are the number of fully open channels in the left and right leads~\cite{nazarov2,Jalabert,Oberholzer}. Working in the limit of large channel number allows us to ignore the effects of weak localization and mesoscopic fluctuations. 

We will find the noise in this case by a direct average of the expression~\cite{bbref,pvb}
\begin{equation}\label{snoise}
S=\frac{e^2}{4h}\int dE\, \mathrm{tr}\{({\cal S}^{\dagger}\Lambda{\cal S}-\Lambda)(\openone+{\cal F}_{E})({\cal S}^{\dagger}\Lambda{\cal S}-\Lambda)(\openone-{\cal F}_{E}) \}
\end{equation}
$\Lambda=(N_L{\cal P}_L-N_R{\cal P}_R)/N^2$, where ${\cal P}_{L/R}$ are projectors onto the channels in the two terminals.  The trace is over the orbital channels in the leads, as well as spin.
An average of Eq.~\ref{snoise} in the $N\gg1$ limit using standard techniques yields~\cite{pvb}.
\begin{eqnarray*}
S=\frac{G_d}{4}\sum_{\alpha\beta, M}\int dE\,({\cal D}\cdot (\openone+{\cal F}_{E\,\alpha}))_M ({\cal D}\cdot (\openone-{\cal F}_{E\,\beta}))_M\\*
 +\frac{e^2}{4h}\int dE\, \mathrm{tr}\{\Lambda(\openone+{\cal F}_{E})\Lambda(\openone-{\cal F}_{E})\}\;.
\end{eqnarray*}
This is the analog of the result (\ref{diff}) for the diffusive wire.  $G_d=(2e^2/h)N_LN_R/N$ is the conductance of the dot. Greek letters index channels in the leads.  ${\cal D}_{LM}$ is the zero-dimensional diffuson given by ${\cal D}_{00}=\langle\mathrm{tr}[{\cal S}_{\alpha\beta}{\cal S}^{\dagger}_{\beta\alpha}]\rangle$, ${\cal D}_{LM}=\langle\mathrm{tr}[\sigma_L{\cal S}_{\alpha\beta}\sigma_M{\cal S}^{\dagger}_{\beta\alpha}]\rangle$ $L,M=1,2,3$, where the trace is over the spin indices. The matrices $\openone\pm{\cal F}$ are understood to be resolved into the singlet-triplet basis as before prior to multiplication by the diffuson.
We will consider two cases: the effect of an external magnetic field only (no spin-orbit coupling), and the effect of spin-orbit coupling only. Following the notation of Ref.~\onlinecite{af}, 
the two cases correspond to, 
%
\begin{eqnarray*}
&&{\cal D}^{(1)}=\left[N+i\epsilon_Z \mathbf{S}\cdot\hat\mathbf{b}\right]^{-1}\\*
&&{\cal D}^{(2)}=\left[N+\epsilon^{\mathrm{SO}}_{\parallel}(S_1^2+S_2^2)+\epsilon^{\mathrm{SO}}_{\perp}S_3^2\right]^{-1}\;.
\end{eqnarray*}
%
%
%
In the above $(S_K)_{LM}=-i\epsilon_{KLM}$ are spin-1 operators.  $\epsilon^{\mathrm{SO}}_{\perp}$ is the rate of spin relaxation in the $x-y$ plane, whereas $\epsilon^{\mathrm{SO}}_{\parallel}$ governs the relaxation of the $z$-component (both these and $\epsilon_Z$ now expressed in units of $\Delta/2\pi\hbar$, where $\Delta$ is the single-particle level-spacing of the closed dot). Straightforward calculation using the distribution function (\ref{bc}) gives the $T=0$ Fano factor
\[F^{(1)}=\frac{N_LN_R}{N^2}+\frac{1}{2}N_L^2\left[\frac{1}{N^2}-\frac{1}{N^2+\epsilon_Z^2}\right]\sin^2\theta\;,\]
in the first case, where $\theta$ is the angle between $\hat\mathbf{s}$ and $\hat{\mathbf{b}}$. 
For the second case, we have
\begin{eqnarray*}
F^{(2)}=\frac{N_LN_R}{N^2}+\frac{1}{2}N_L^2\left[\frac{1}{N^2}-\cos^2\theta_z\frac{1}{(N+2\epsilon_{\parallel}^{\mathrm{SO}})^2}\right.\\*-\left.\sin^2\theta_z\frac{1}{(N+\epsilon_{\parallel}^{\mathrm{SO}}+\epsilon_{\perp}^{\mathrm{SO}})^2}\right]\;,
\end{eqnarray*}
where $\theta_z$ is the angle to the $z$-axis. A feature of this two-dimensional system is that, since $\epsilon_{\perp}^{\mathrm{SO}}\gtrsim\epsilon_{\parallel}^{\mathrm{SO}}$ the deviation of the Fano factor from $N_LN_R/N^2$ depends on the direction of polarization.  In all cases the Fano factor of a symmetrical dot ($N_L=N_R$) increases from $1/4$ to a maximum value of $3/8$ with increasing spin-flip rate.  

\begin{acknowledgments} 
This research is supported in part by the David and Lucille Packard foundation, and by the National Science Foundation under Grant No. PHY99-07949.
\end{acknowledgments}



\begin{thebibliography}{99}
\bibitem{foot1} The definition of the noise used here - a factor of 2 smaller than the convention used in e.g. Ref.~\onlinecite{bbref} - is the natural one in the context of FCS.
\end{thebibliography}
\end{document}